# Prospects for Quantum Computing

Subhash Kak

**Introduction**

The idea of quantum computing is attractive. It enriches computing theory beyond classical techniques in doing Fourier transform faster (albeit the solution is not fully available), factorization in *log n* rather than $\sqrt{n}$ steps, search for an item with a specific property in an unordered database in $\sqrt{n}$ rather than *n/2* steps, solving Pell's equation in polynomial time, and it has applications to cryptography [1]. In theory, quantum computing can solve problems where the solution amounts a rotation in an appropriate space (as in primality testing since integers form a multiplicative group). The allure of finding new problems that could be shown to be solved faster by quantum algorithms and the challenge of building quantum computing machines has made the field popular amongst mathematicians and physicists. It has spurred research in materials science and in the implementation of quantum circuits.

Soon after proposals for quantum computing were advanced, it was argued that problems of decoherence and noise precluded effective physical implementation [2]. The computing system is not to interact with the environment lest it decohere and at the same time it is to strongly and precisely interact with the control circuitry!

It is supposed that useful computations may be performed within the time the system is coherent. Quantum error correcting codes that are a generalization of classical error correcting codes have been proposed to meet the problem of random errors. Their effectiveness is predicated on certain assumptions regarding noise. Quantum error correction codes consider only two kinds of errors, namely qubit and phase flips, and it is assumed that at least some of the gates are ideal.

In the classical world there are two steps to error correction: first, "small errors" are corrected because the ideal representation of the bits is known and small departures from it may be corrected readily using non-linear thresholding operations; second, "large errors", which are the infrequent bit flips, may be correcting by redundancy coding. Quantum error correction schemes do not have the capacity to perform the ubiquitous and essential nonlinear corrections [3],[4].

Researchers in quantum error correction invoke the so-called "threshold theorem", according to which if the error per qubit per gate is below a certain value then under certain conditions computation with some qubits could take place reliably [5], [6], [7]. Experimental evidence to support the assumption that error rate for physical implementation of gates will be below this threshold is not yet available.

Fault-tolerant conceptual schemes assume that a system could be built recursively (e.g. [7], page 40) where each noisy gate is replaced by an ideal gate by the use of error correction circuitry. A non-recursive, so-called Fibonacci scheme is described in a recent paper [8]. It assumes the existence of perfect gates in the use of teleportation for error correction. It also uses Bell states to generate which we need perfect gates. The scheme assumes that there are no gates with leakage or correlated faults.

Quantum error correction framework is reasonable in the context of communication, where the qubits are known (as a generalization of error correction in classical communication) since this involves a single stage of induced error in the qubits. Classical error correction framework also applies to correction in computing since the bit blocks, which may be unknown in a specific computation, are not associated with any continuous parameter of significance to the computation process. If the errors are small, redundancy is enough to take the computed bits to the correct unknown sequence.

This framework does not quite apply to the quantum circuit model of computation if the qubits in intermediate stages suffer random and continuous phase errors in addition to bit flips and phase flips. If the continuous phase errors are unknown they cannot be corrected. One cannot assume that error correction can be carried out in only one stage at the very end of a long chain of quantum gates, because the errors in the intermediate stages could become catastrophically large. Dyakonov presents a critical evaluation [9] of fault-tolerant quantum computing and of the "threshold theorem". According to this view, safeguarding qubits, without safeguarding the phase that can suffer continuous errors, makes fault tolerance in quantum computing impractical.

In this note, we review first the question of realizability of quantum gates. Physical gates are bound to have some errors, and as non-ideal gates they need to be tested for a wide range of probability amplitudes. Experimental implementations of the controlled-NOT gate have non-unitarity and residual errors. We present some relevant observations on quantum information and cryptography.

**Testing a Quantum Gate**

DeMarco et al [10] present the best-known implementation of the controlled-NOT gate in which the qubits are the spin and the internal energy (ground and second excited) states of a $^9Be^+$ ion. Table 1 presents the performance and as we can see the probabilities do not sum to unity because these data represent the results of four separate experiments. The fact that checking the performance of the gate in this limited setting itself requires different experiments with different control conditions is symptomatic of the difficulty of testing and it shows that the physical implementation cannot be taken to be an ideal gate.

Table 1: Experimental results of controlled-NOT gate [10]

|     | ↓ | ↑ |
| --- | --- | --- |
| n=0 | 0.989 ± 0.006 | 0.050 ± 0.007 |
| n=2 | 0.019 ± 0.007 | 0.968 ± 0.007 |

The authors mention errors in the initial state preparation and "limited gate fidelity" as contributing to the less than perfect performance of the gate. For ease of comparison with the standard gate transformation, one may rewrite the results of Table 1 by representing the internal energy and the spin states in terms of the same qubits as follows:

$$|00\rangle \xrightarrow{CNOT} 0.989|00\rangle + 0.050|10\rangle$$
$$|10\rangle \xrightarrow{CNOT} 0.019|00\rangle + 0.968|11\rangle$$

which may be idealized to

$$|00\rangle \xrightarrow{CNOT} |00\rangle$$
$$|10\rangle \xrightarrow{CNOT} |11\rangle$$

Since we don't expect the idealized transformation to be achieved in practice, testing the gate for only two points for a transformation whose range (in terms of the probability amplitudes) is continuous over complex values in the range (0,1) is unsatisfactory. This is like testing the transformation $y=x^2$ for two values of $x$ and taking that to be true for all values. Two non-ideal gates that give the same value for two points may yet have different characteristics for other points. The effect of the *controlling circuitry cannot be taken to be the same for all values of the probability amplitudes*.

In another implementation of the controlled-NOT gate, based on a string of trapped ions [11] whose electronic states represent the qubits where the control is exercised by focused laser beams, the fidelity of the gate operation *was in the range 70 to 80%* [12]. These were ascribed to laser frequency noise, intensity fluctuations, detuning error, residual thermal excitation, addressing error, and off-resonant excitation. Clearly, this method is not a promising candidate for creating an accurate controlled-NOT gate.

For a recent (c. 2008) optical fiber quantum controlled-NOT gate [13], the authors claim an average logical fidelity of 90% and process fidelity in the range 0.83 to 0.91. The estimated second range is broader because of the substantial errors in the photon sources. The performance is far from ideal.

It is important to note that although the very name of the gate is controlled-NOT, the transformation is not that of one input controlling the other in all cases of the input qubits. As example, if the input state is $\frac{1}{2}(|0\rangle+|1\rangle)(|0\rangle-|1\rangle)$ the output state is $\frac{1}{2}(|0\rangle-|1\rangle)(|0\rangle-|1\rangle)$, in which the first qubit has been transformed under the influence of the second qubit rather than the other way around.

For a *non-ideal* controlled-NOT gate, we should be able to show that:

$$a|00\rangle + b|01\rangle + c|10\rangle + d|11\rangle \xrightarrow{CNOT} a|00\rangle + b|01\rangle + c|11\rangle + d|10\rangle$$

within certain limits of error. In other words, the complex amplitudes (a, b, c, d) should map into (a, b, d, c) with some small error. It is true if the gate was ideal then testing beyond the two cases of 10 and 11 would not be necessary. But since any physical implementation of the gate would not be ideal, one needs to check the working for the entire range of probability amplitudes to ensure that the errors are within the acceptable threshold.

The testing of the physical gate requires that various (a, b, c, d) amplitudes be generated with perfect fidelity, which requires that a perfectly functioning two-qubit gate (that may be controlled-NOT gate) should already exist. Since each of the values (a, b, c, d) is complex and over (0,1) and testing causes collapse to the components along the measurement bases, certification that a quantum gate works within a specific error performance will require considerable testing.

To test a non-ideal controlled-NOT gate, we first need a perfectly functioning controlled-NOT gate to generate complex superpositions. Current experimental evidence on the implementation of the controlled-NOT gate indicates non-unitarity and substantial errors.

**Quantum Information**

We now consider the question of information in a qubit. In communication theory, information is a property that is associated with the surprise the received message has for the recipient. Classical entropy is given by $H(X) = -\sum_x p_x \log p_x$, where $p_x$ is the probability of the message *x*.

Quantum information is traditionally measured by the von Neumann entropy, $S_n(\rho) = -tr(\rho \log \rho)$, where $\rho$ is the density operator associated with the state. This entropy may be written also as

$$S_n(\rho) = -\sum_x \lambda_x \log \lambda_x$$

where $\lambda_x$ are the eigenvalues of $\rho$. It follows that a pure state has zero von Neumann entropy (we will take the *log* in the expression to the base 2 implying the units are *bits*). The von Neumann measure is a thermodynamic measure and it is independent of the receiver.

Consider an entangled pair of quantum objects represented by the pure state
$|\Psi\rangle = \frac{1}{\sqrt{2}}(|00\rangle + |11\rangle)$ that corresponds to the density operator $\rho = \begin{bmatrix} .5 & 0 & 0 & .5 \\ 0 & 0 & 0 & 0 \\ 0 & 0 & 0 & 0 \\ .5 & 0 & 0 & .5 \end{bmatrix}$. Its
eigenvalues are 0, 0, 0, and 1, and, therefore, its von Neumann entropy is zero. But consider the two objects separately; their density operators are $\begin{bmatrix} .5 & 0 \\ 0 & .5 \end{bmatrix}$ each, which means that their entropy is 1 bit. We are confronted with the situation that two components of a quantum system each have non-zero entropy although the system taken as a whole has zero entropy! This is not intuitively reasonable.

The concept of information in classical theory is related to the reduction in uncertainty by the received communication. From this perspective, an unknown pure state does communicate information to the recipient of the state [13]. For an unknown pure state $|\Psi\rangle = a|0\rangle + b|1\rangle$, the many copies of it that are received help the recipient estimate the values of $a$ and $b$. If the objective is the estimation of the density matrix $\rho$, we can do so by finding average values of the observables $\rho$, $X\rho$, $Y\rho$, $Z\rho$ by means of the expansion:

$$\rho = \frac{1}{2}(tr(\rho)I + tr(X\rho)X + tr(Y\rho)Y + tr(Z\rho)Z)$$

Considering a more constrained setting, assume that the information that user Alice, who is preparing the states, is trying to communicate to the user Bob, is in the ratio $a/b$, expanded as a numerical sequence, $m$. For further simplicity it is assumed that Alice and Bob have agreed that $a$ and $b$ are real. Then $a = \frac{m}{\sqrt{1+m^2}}$ and $b = \pm\frac{1}{\sqrt{1+m^2}}$. One may use either a pure state $|\Psi\rangle = a|0\rangle + b|1\rangle$ or a mixed state consisting of an equal mixture he the states $|\varphi_1\rangle = a|0\rangle + b|1\rangle$ and $|\varphi_2\rangle = a|0\rangle - b|1\rangle$. Alternatively, one may communicate the ratio $a^2/b^2 = m$. In this case, $a = \frac{\sqrt{m}}{\sqrt{1+m}}$ and $b = \pm\frac{1}{\sqrt{1+m}}$. Since $b^2 = \frac{1}{1+m}$, one needs to merely determine the sequence corresponding to the reciprocal of $1+m$ for the probability of the component state $|1\rangle$.

The idea of zero entropy for an unknown pure state is reasonable from the perspective that once the state has been identified; there is no further information to be gained from examining its copies. But it is not reasonable if the game being played between the sender and the receiver consists of the sender choosing one out of a certain number of polarization states (say, for a photon) and supplying several copies of it to the receiver. In this latter case, the measurements made on the copies do reveal information regarding the choice made. If the set of choices is infinite, then the "information" generated by the source is unbounded. From the point of view of the preparer of the states, the information in the pure state is limited by the "relationship" between the source and the receiver, and by the precision of the receiver's measurement apparatus. If the sender chose a polarization state that the receiver's measurement apparatus was already synchronized with, the receiver could recognize the state quite readily.

Sometime ago, I proposed a measure of quantum entropy that covers both pure and mixed states [14]. It has two components: one informational (related to the pure components of the quantum state, which can vary from receiver to receiver), and the other that is thermodynamic (which is receiver independent). For a two-component elementary mixed state, the most information in each measurement is one bit, and each further measurement of identically prepared states is also one bit. For an unknown pure state, the information in it represents the choice the source has made out of the infinity of choices related to the values of the probability amplitudes with respect to the basis components of the receiver's measurement apparatus. Each measurement of a two-component pure state will provide at most one bit of information, and if the source has made available an unlimited number of identically prepared states the receiver can obtain additional information from each measurement until the probability amplitudes have been correctly estimated. Once that has occurred, unlike the case of a mixed state, no further information will be obtained from testing additional copies of this pure state.

The receiver can make his estimate by adjusting the basis vectors so that he gets *closer* to the unknown pure state. The information that can be obtained from such a state in repeated experiments is potentially infinite in the most general case. But if the observer is told what the pure state is, the information associated with the states vanishes, suggesting that a fundamental divide exists between objective and subjective information.

We believe consideration of the problem of quantum information is important to applications that deal with transactional aspects of information in computing and cryptography.

**Concluding Remarks**

Current progress in physical implementations is limited and there is no unanimity on what kind of physical qubits to use [10]. DiVincenzo [15] has listed five requirements for successful physical implementation and a qubit system that would meet each of these requirements in an easy manner remains to be identified.

We have argued that the von Neumann measure of quantum information does not cover all aspects of transactional – as against structural – information.

Quantum cryptography has been physically demonstrated and it has reached the marketplace, but it faces challenges to be competitive with classical cryptography. The main advantage of the BB84 protocol over classical cryptography is that eavesdropping can be detected in theory. In practice, lasers that are used for the implementation of BB84 do not transmit single photons [16] and, therefore, BB84 is susceptible to siphoning attack. The three-stage quantum cryptography protocol [17] is not susceptible to the siphoning attack, but this is at the cost of sending qubits over three hops.